\documentclass[conference]{IEEEtran}
\IEEEoverridecommandlockouts

\usepackage{cite}
\usepackage{amsmath,amssymb,amsfonts}
\usepackage{algorithmic}
\usepackage{graphicx}
\usepackage{CJKutf8}
\usepackage{textcomp}
\usepackage{xcolor}
\usepackage{multirow}
\def\BibTeX{{\rm B\kern-.05em{\sc i\kern-.025em b}\kern-.08em
    T\kern-.1667em\lower.7ex\hbox{E}\kern-.125emX}}

\title{Automatic Text Pronunciation Correlation Generation and Application for Contextual Biasing\\
\thanks{\textsuperscript{†}Corresponding author.}
\thanks{This work is partially supported by the National Natural Science Foundation of China (No.62401560), the Youth Innovation Promotion Association Chinese Academy of Sciences, the Basic and Frontier Exploration Project Independently Deployed by Institute of Acoustics, Chinese Academy of Sciences (No.JCQY202411).}
}

\author{
	\IEEEauthorblockN{Gaofeng Cheng$^{1,2,\textit{\dag}}$, Haitian Lu$^{1,2}$, Chengxu Yang$^{1,2}$, Xuyang Wang$^{1}$,Ta Li$^{1,2}$, Yonghong Yan$^{1,2}$}
	\IEEEauthorblockA{
		$^1$ Key Laboratory of Speech Acoustics and Content Understanding, Institute of Acoustics, CAS, China \\
		$^2$ University of Chinese Academy of Sciences, China \\
        \IEEEauthorblockA{\{chenggaofeng, luhaitian, yangchengxu, wangxuyang, lita, yanyonghong\}@hccl.ioa.ac.cn \\
	}
	}
        }

\begin{document}
\fontsize{9pt}{11pt}\selectfont
\begin{CJK}{UTF8}{gbsn}

\maketitle

\begin{abstract}
Effectively distinguishing the pronunciation correlations between different written texts is a significant issue in linguistic acoustics. Traditionally, such pronunciation correlations are obtained through manually designed pronunciation lexicons. In this paper, we propose a data-driven method to automatically acquire these pronunciation correlations, called automatic text pronunciation correlation (ATPC). The supervision required for this method is consistent with the supervision needed for training end-to-end automatic speech recognition (E2E-ASR) systems, i.e., speech and corresponding text annotations. First, the iteratively-trained timestamp estimator (ITSE) algorithm is employed to align the speech with their corresponding annotated text symbols. Then, a speech encoder is used to convert the speech into speech embeddings. Finally, we compare the speech embeddings distances of different text symbols to obtain ATPC. Experimental results on Mandarin show that ATPC enhances E2E-ASR performance in contextual biasing and holds promise for dialects or languages lacking artificial pronunciation lexicons. 
\end{abstract}


\section{Introduction}
Pronunciation plays a crucial role in linguistic acoustics, especially when dealing with languages that exhibit a high degree of phonetic variability. The ability to effectively distinguish pronunciation correlations between different written texts is important for improving various language processing tasks, such as automatic speech recognition (ASR) and text-to-speech (TTS) \cite{intro-asr-lexi1-jstsp,intro-tts-lexi1-taslp,ee-kws,kaiyu-modular-ee-asr}. Traditionally, these pronunciation correlations have been obtained through the manual design of pronunciation lexicons. However, this approach is inherently labor-intensive, language-specific \cite{davel2009pronunciation,pronun-for-asr}.

The advent of E2E-ASR models has brought about significant advancements in modeling the speech-to-text directly  without relying on handcrafted features or intermediate representations \cite{li2021recent,eteh-ee-asr,las-ee-asr,online-ee-asr}. However, the vanilla EE-ASR architectures fail to model the text-to-text pronunciation correlations effectively, the accurate modeling of pronunciation correlations remains a challenge. This is particularly true in scenarios where the pronunciation of certain text symbols can vary depending on dialect, or the presence of specialized vocabulary. 
This shortcoming poses significant challenges when implementing biasing recognition \cite{bruguier2019phoebe,Towards-Contextual-Spelling-Correction-ee-asr}, which always requires a more nuanced understanding of text pronunciation correlation.

In this paper, we introduce a novel data-driven method for automatically acquiring pronunciation correlations, termed automatic text pronunciation correlation (ATPC). Without relying on manually designed pronunciation lexicons, ATPC leverages the same supervision as E2E-ASR systems, utilizing pairs of speech and their corresponding text annotations.
The ATPC generation procedure consists of three main stages:\textbf{ 1. text-speech alignment; 2. speech embedding extraction and segmentation; 3. pronunciation correlation calculation.}
The generation procedure begins by employing the ITSE algorithm  \cite{itse} to align text symbols with the corresponding speech. Then we use a speech encoder to convert the raw speech into speech embeddings. We explored several ways to make the speech embeddings more distinguishable. Finally, the core of the ATPC method lies in the comparison of speech embedding distances between different text symbols. By analyzing these distances, our method can automatically derive pronunciation correlations, effectively distinguishing between subtle variations in pronunciation across different texts. 

The ATPC has multiple potential applications, with one prominent use being its enhancement of E2E-ASR systems through contextual biasing—where certain keywords or phrases are given higher recognition priority. 
We validate the effectiveness of ATPC through a series of experiments conducted on Mandarin\footnote{The Mandarin ATPC matrix will continue updating at https://github.com/\\SpeechClub/ATPC/blob/main/aishell2-mandarin-atpc.txt}, a language known for its tonal variations and complex phonetic structure. The results demonstrate that ATPC not only effectively distinguishes pronunciation distances between different texts but also serves as a useful plug-in module for E2E-ASR models. The integration of ATPC into E2E-ASR enhances its ability to handle contextual biasing, leading to more accurate and reliable speech recognition outcomes.

In general, our work presents a step forward in the automatic modeling of pronunciation correlations, offering a learnable solution that can be readily integrated into modern E2E-ASR systems. The ATPC method represents a promising direction for future research in linguistic acoustics.

\begin{figure*}[ht]
    \centering
    \includegraphics[width=0.95\textwidth]{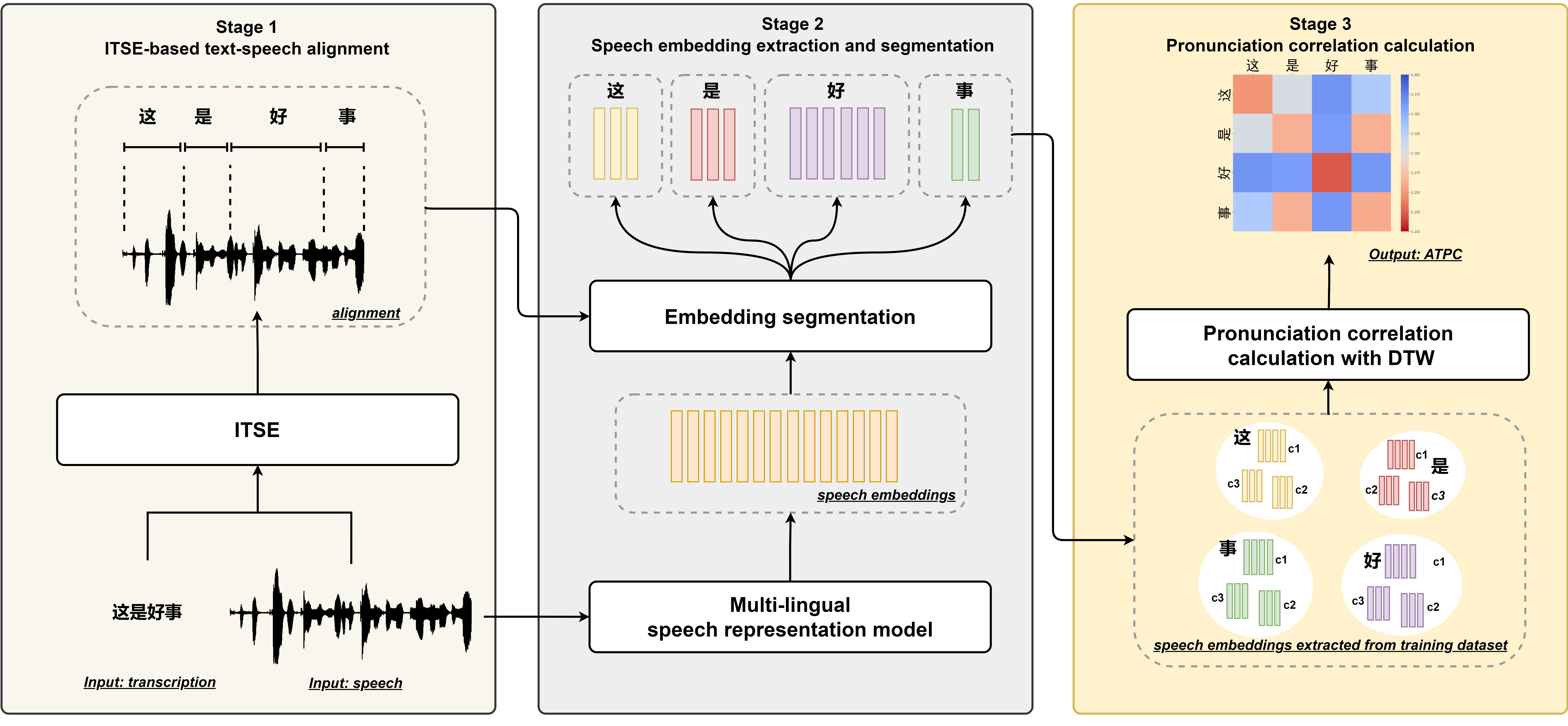}
    \caption{The overall diagram of generating ATPC, with c1, c2, and c3 represent multiple embeddings corresponding to the same character in the training dataset.}
    \label{fig:diagram}
\end{figure*}

The rest of this paper is organized as follows. In Section \ref{sec:awpc_generate}, we describe how we generate the ATPC on Mandarin. In Section \ref{sec:exp-setup}, we describe the experimental setups we adopt. Experimental results are described in Section \ref{sec:results}. And we present conclusion and future work in Section \ref{sec:conclusion}.

\section{Generation Procedure of the Proposed ATPC}
\label{sec:awpc_generate}

As depicted in Figure \ref{fig:diagram}, the generation of ATPC comprises three principal stages. Initially, ITSE-based text-speech alignment is employed to determine the onset and offset timestamps for each character. Subsequently, speech embeddings are extracted utilizing multi-lingual speech representation models, with segmentation aligned to the derived timestamps. Lastly, the pronunciation correlation between distinct characters is calculated, culminating in the generation of ATPC. This entire process is conducted without reliance on specialized knowledge such as a pronunciation lexicon.


\subsection{ITSE-based Text-speech Alignment}
\label{subsec:ta-as}

As illustrated in the left block of Figure \ref{fig:diagram}, ITSE is employed to conduct text-to-speech alignment. In contrast to CTC-based segmentation, ITSE provides both the start and end accurate timestamps for each character. In contrast to traditional GMM-HMM-based forced alignment, ITSE operates without the need for pronunciation lexicons.

Leveraging E2E-ASR, ITSE facilitates token-level text-to-speech alignment. The training of ITSE begins with coarse initial alignment targets derived from connectionist temporal classification (CTC) \cite{graves2006ctc} posteriors. Throughout the training process, iterative realignment is conducted to refine and update the targets. We use Mandarin characters as the target token of ITSE. A detailed explanation of ITSE can be found in \cite{itse}.

\subsection{Speech Embedding Extraction and Segmentation}
\label{subsec:embedding}

As illustrated in the middle block of Figure \ref{fig:diagram}, to represent the pronunciation patterns of different Mandarin characters, we use multi-lingual speech representation models to extract speech embeddings. We opt for embedding segmentation rather than audio segmentation because the extraction of speech embeddings relies on contextual information. Firstly, we extract speech embeddings on the entire utterance. Secondly, we segment the extracted embeddings according to the text-to-speech alignment results.


For speech embedding extraction, different layers in the speech representation models encode different information, where the shallowest layers encode acoustic features, followed by phonetic, word identity, and word meaning information \cite{pasad2021layer}. We experiment with different multi-lingual speech representation models and different layers to obtain better pronunciation distinguishing ability. The results are presented in Section \ref{subsec: distinguish}. It is worth noting that the training of the multilingual speech representation models we use does not rely on manually crafted pronunciation lexicons. 

For embedding segmentation, speech embeddings are extracted at a frequency of $F$ Hz (every $1000/F$ ms). The segmentation intervals are determined by dividing the start and end timestamps of each character by $1000/F$ ms and rounding the resulting values.

\subsection{Text Pronunciation Correlation Calculation}
\label{subsec:correlation}

As illustrated in the right block of Figure \ref{fig:diagram}, the dynamic time warping (DTW) algorithm \cite{dtw} is employed to calculate pronunciation correlation. DTW effectively measures the distance between two embeddings of differing lengths while preserving the sequential integrity of phonetic elements in Mandarin characters. Firstly, we construct the embedding set for Mandarin characters. Secondly, we calculate the pronunciation correlation between each two characters and generate the ATPC matrix.

For the construction of the embedding set, we randomly select $E$ embeddings for each character in the training dataset. If a character occurs fewer than $T$ times, all available embeddings are included. Characters with fewer than 3 occurrences are excluded to avoid unreliable correlations due to data scarcity.

\begin{figure}[t]
    \centering
    \includegraphics[width=0.32\textwidth]{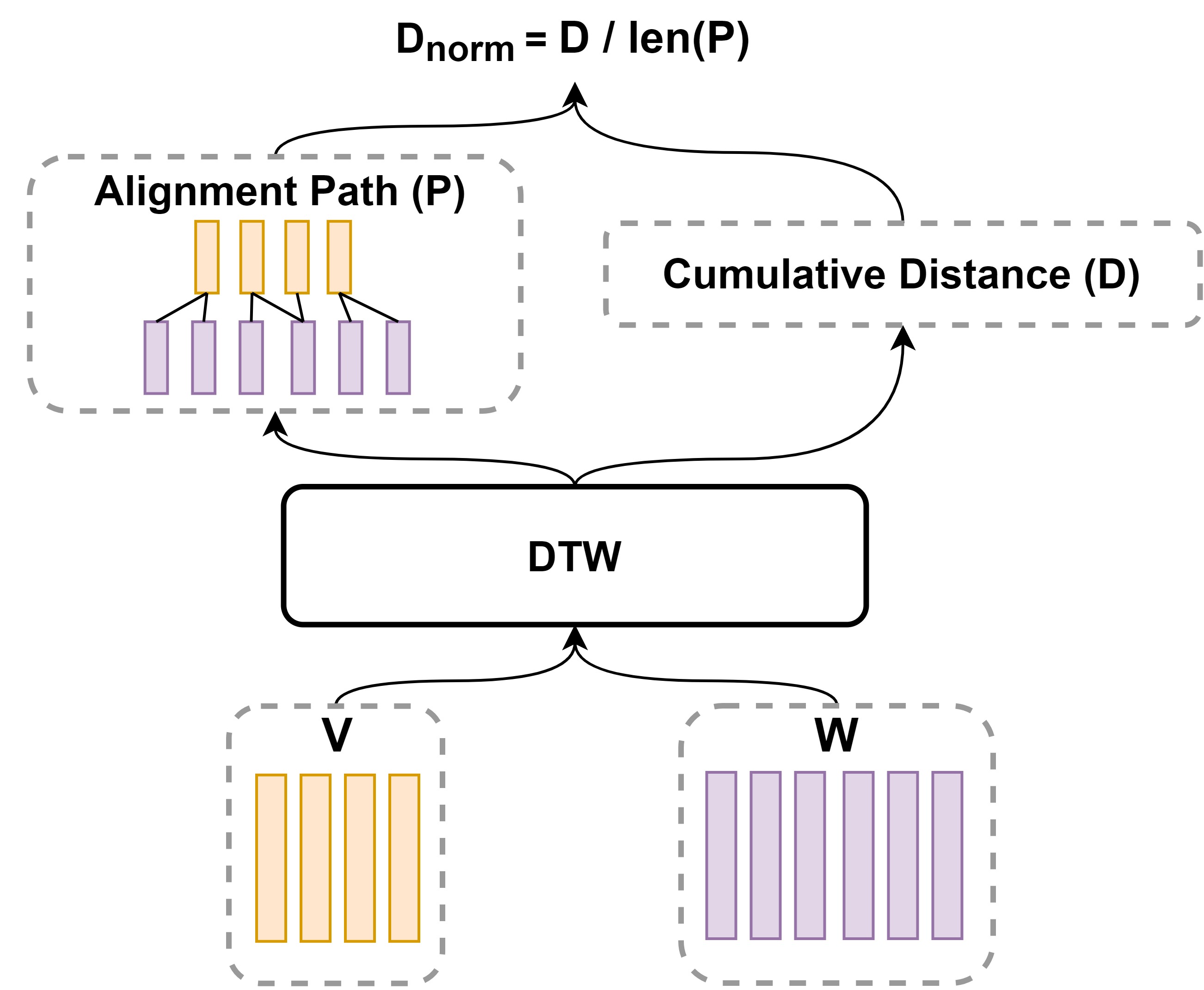}
    \caption{Pronunciation correlation calculation with DTW. V and W represent the speech embeddings of two different Mandarin characters. $D_{norm}$ is the pronunciation correlation between V and W. The alignment path is obtained by tracing backward through the DTW table, iteratively choosing the previous points with the lowest cumulative distance.}
    \label{fig:dtw}
\end{figure}

For ATPC calculation, we apply DTW as depicted in Figure \ref{fig:dtw}, to each two characters in the embedding set. We compute $D_{\text{norm}}$ for all pairs of embeddings of the two characters and calculate the average distance as Equation \ref{equation: 1}. 

\begin{equation}
\label{equation: 1}
\text{Dist}(c_j, c_k) = \frac{1}{M \times N} \sum_{m=1}^{M} \sum_{n=1}^{N} \text{DTW}(V_j^m, W_k^n)
\end{equation}

In Equation \ref{equation: 1}, $c_j$ and $c_k$ represent the $j$th character and the $k$th character in the training dataset. $M$ and $N$ are the number of embeddings for $c_j$ and $c_k$, respectively. $V_j^m$ and $W_k^n$ are the $m$th element and the $n$th element of the embedding set for $c_j$ and $c_k$, respectively. $\text{Dist}(c_j, c_k)$ is the element in the $j$th row and $k$th column of the ATPC matrix.

\section{Experimental Setup}
\label{sec:exp-setup}

\subsection{Datasets}

To train the speech representation model, we use a subset of BABEL from IARPA \cite{babel}, which is is a multilingual  conversational telephone speech corpus. The subset includes 23 languages: Cantonese, Assamese, Bengali, Pashto, Turkish, Tagalog, Vietnamese, Haitian Creole, Swahili, Lao, Tamil, Kurmanji Kurdish, Zulu, Tok Pisin, Cebuano, Kazakh, Telugu, Lithuanian, Paraguayan Guarani, Igbo, Amharic, Dholuo, and Georgian. 

To better validate the cross-lingual performance of the proposed ATPC generation method, we use the training set of Aishell-2 \cite{du2018aishell2}, a Mandarin speech corpus, to train ITSE and generate ATPC.

To evaluate the effectiveness of ATPC in the context of the contextual biasing, we employed the Aishell-1 \cite{bu2017aishell1} contextual biasing dataset\footnote{The hotword list can be found at https://github.com/SpeechClub/ATPC/\\blob/main/aishell1-hotword161.txt}. The development set comprises 1334 sentences, featuring a total of 600 hotwords while the test set comprises 235 sentences, featuring a total of 161 hotwords, with each sentence containing at least one hotword.

\subsection{Evaluation Metrics}

To evaluate the pronunciation distinguishing ability of different speech embeddings, we calculate the DTW distance between homophones and non-homophones in Mandarin. Additionally, we calculate the relative disparity between the DTW distances of homophones and non-homophones. A greater relative disparity indicates better distinguishing performance of the speech embedding.

In addition to the conventional Character Error Rate (CER), our evaluation metrics include a Biased Character Error Rate (B-CER) and an Unbiased Character Error Rate (U-CER), as referenced in \cite{huang2023contextualized}. 
Additionally, we will calculate the recall, precision, and F1 score (R/P/F) across all hotwords to provide a comprehensive assessment of performance.

\subsection{ATPC Generation}
\label{subsec:atpc-generation}

We use XLSR-53 as the backbone model \cite{conneau21_interspeech}, and further finetune it on Babel IPA recognition task. Firstly, we convert the transcripts of the BABEL training set into IPA sequences using the pronunciation lexicon of the dataset. Secondly, we employ the IPA sequences as the training target and finetune the backbone model on speech recognition task. It is worth noting that we did not use the pronunciation lexicon of Mandarin because the subset of BABEL dataset we use does not contain Mandarin speech and the training of XLSR-53 is self-supervised.

We also experiment with embeddings from different layers of the speech representation model and different distance functions for the DTW algorithm. 

For speech embedding extraction in Section \ref{subsec:embedding}, embeddings are extracted at a frequency of 50Hz (every 20ms). For the construction of the embedding set in Section \ref{subsec:correlation}, we select 100 embeddings for each character and delete the characters occur fewer than 3 times in the training set of Aishell-2. We obtain embeddings of 3711 characters in total and generate the ATPC, which is a $3711 \times 3711$ matrix representing the distances between each two Mandarin characters.

\subsection{Contextual Biasing}

We utilized Wenet\cite{zhang2022wenet}, an open-source E2E-ASR framework, as the basis for our contextual biasing experiments. The baseline model we chose is a pre-trained checkpoint of Wenet's conformer architecture, specifically trained on the Aishell-1 dataset. This model features 12 conformer layers within its encoder and 6 bi-transformer layers in the decoder, both operating with 256-dimensional inputs and incorporating 4 self-attention heads. 

For utilizing ATPC matrix for contextual biasing, we perform row-wise normalization, ensuring that the diagonal elements are set to 1.0. This normalization implies that any value less than 1.0 signifies a shorter distance between the respective pair of characters, thereby indicating a stronger association or similarity between them. Following the normalization of the ATPC matrix, we identify candidate replacement characters for each character in the ASR decoding result by selecting those with a distance of less than 1.07. 
To determine the threshold of 1.07, we conduct hot word biasing experiments on the Aishell-1 contextual biasing development set. We evaluate thresholds ranging from 1.01 to 1.09, where a threshold of 1.07 achieves the lowest CER and the second-highest F1 score. During the decoding process, we match each word against potential hotword replacements and record the average distance for these matches. Finally, we proceed with hotword replacements in ascending order of distance, starting with the closest matches.

We utilize a WFST-based context decoding graph method, which is a kind of shallow-fusion contextual biasing methods \cite{zhao2019shallow,kim2021improved-shallow-fusion}, as our baseline, with the context graph (C-g) score configured at 6.0. For the deep contextual biasing baseline \cite{han2021cif-deep-fusion,chen2019joint-deep-fusion}, we adopt the contextual phrase prediction network based on AED-CTC structure (CPPN), as detailed in \cite{huang2023contextualized}. The contextual phrases lists of CPPN and ATPC are identical. This network architecture includes a context encoder, a biasing layer, a context decoder, and CTC loss. The context encoder comprises two layers of BLSTM and a subsequent linear layer. The biasing layer features 4-head attention layers and an additional linear layer. The context decoder incorporates a linear layer that maps the input dimension to the vocabulary size. We set the deep biasing score to 2.0. The training process for this setup aligns with the methodology described in \cite{huang2023contextualized}.

\section{Results}
\label{sec:results}

\begin{figure}[ht]
    \centering
    \includegraphics[width=0.4\textwidth]{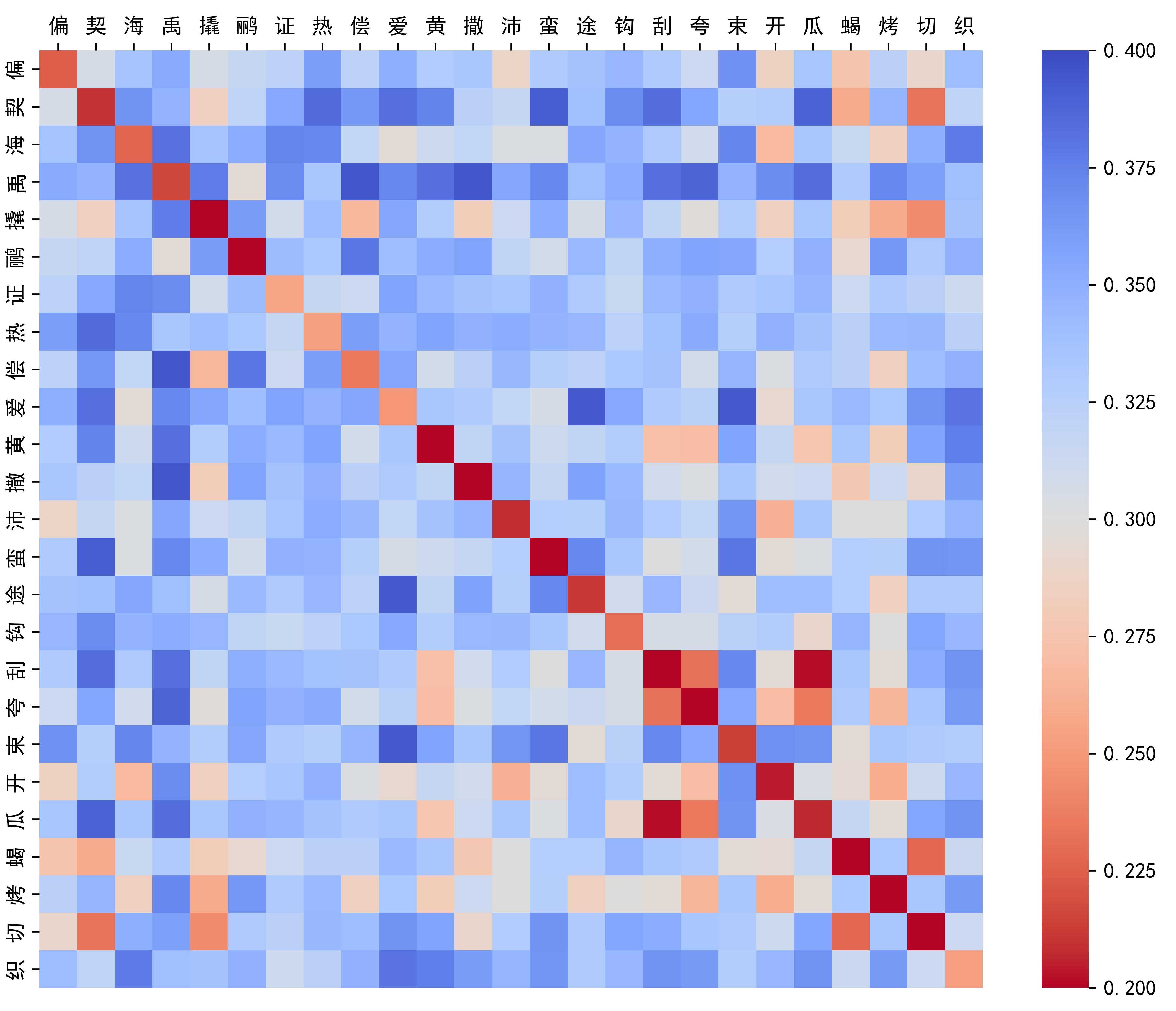}
    \caption{The visual analysis of generated ATPC matrix subset.}
    \label{fig:atpc}
\end{figure}

\subsection{Pronunciation Distinguishing Ability of Different Embeddings}
\label{subsec: distinguish}

As shown in the results in Table \ref{table:distinguish}, speech representation models finetuned on IPA recognition consistently outperform the backbone model XLSR-53 on distinguishing pronunciation. For instance, the relative disparity between the DTW distances of homophones and non-homophones increases from 19.7\% to 21.1\% when comparing XLSR-layer15 to IPA-layer15 using Euclidean distance. However, the embeddings from different layers do not consistently exhibit variations in their distinguishing capabilities, while the embeddings of layer 15 achieve the best performance in most cases. As for the distance function, cosine distance consistently outperforms the Euclidean distance. For example, the relative disparity increases from 21.1\% to 28.8\% for IPA-layer15 with Euclidean and cosine distance.

According to the analysis above, we select the 15th-layer embedding from the multi-lingual speech representation model fine-tuned for IPA recognition, along with cosine distance as the metric in our experiments.

\begin{table}[t]
\centering
\caption{Average DTW distance of homophones and non-homophones with Euclidean distance and Cosine distance. IPA represents the XLSR-53 model finetuned on the Babel IPA task. Distance Rel Disparity is the distance gap between homophones and non-homophones.}
\renewcommand{\arraystretch}{1.5}
\setlength{\tabcolsep}{2mm}{
\begin{tabular}{c|c|c|c}
\hline
\multirow{3}{*}{Model} & \multicolumn{3}{c}{Euclidean / Cosine Distance} \\ \cline{2-4} 
& \multirow{2}{*}{Homophones} & \multirow{2}{*}{Non-homophones} & \multirow{2}{*}{\shortstack{Distance Rel\\ Disparity} $\uparrow$} \\
&&&\\
\hline\hline
XLSR-layer12 & 112.90 / 0.247 & 127.79 / 0.314 & 11.7\% / 21.3\% \\ \hline
XLSR-layer15 & 105.67 / 0.183 & 131.66 / 0.258 & 19.7\% / 29.1\% \\ \hline
XLSR-layer18 & 145.54 / 0.197 & 167.08 / 0.255 & 12.9\% / 22.7\% \\ \hline
IPA-layer12 & 419.51 / 0.174 & 491.23 / 0.224  & 14.6\% / 22.3\% \\ \hline
IPA-layer15 & 394.47 / 0.136 & 499.87 / 0.191 & \textbf{21.1\%} / 28.8\% \\ \hline
IPA-layer18 & 495.55 / 0.122 & 588.40 / 0.189 & 15.8\% / \textbf{35.4\%} \\ \hline
\end{tabular}
}
\label{table:distinguish}
\end{table}

\subsection{Visual Analysis of ATPC matrix on Mandarin}

A subset of ATPC matrix is visualized in Figure \ref{fig:atpc}. Each value in the matrix represents the DTW distance between corresponding Mandarin characters. As is shown in Figure \ref{fig:atpc}, the DTW distance between ``刮" (gua1) and ``瓜" (gua1) is relatively low, reflecting their homophonic nature in Mandarin, whereas the DTW distance between ``爱" (ai4) and ``途" (tu2) is significantly higher, aligning with expectations since these characters share no common phones. We can conclude from the visual analysis above that the ATPC matrix generally reflects the pronunciation correlation between Mandarin characters. 

\subsection{Contextual Biasing Effectiveness of ATPC}
To objectively evaluate the effectiveness of ATPC, we selected a Mandarin speech corpus. It is easy to find a standard manually crafted pronunciation lexicon suitable for this corpus.

The experimental results of the contextual biasing task are shown in Tabel \ref{tab:contexual biasing result}. 
Compared to the baseline, the implementation of ATPC has resulted in a notable improvement across several metrics. We observed a relative reduction of 13.0\% in CER and an even more significant average relative decrease of 22.5\% in B-CER (Row 1 versus Row 4). Additionally, the recall rate for hotwords has increased by 25\%, and the F1 score has seen a substantial improvement, rising by 24\% (Row 1 versus Row 4). 

Compared with the CPPN based deep biasing, the ATPC consistently outperforms, moreover, ATPC does not necessitate the incorporation of additional neural networks that require separate training, thereby preserving the original E2E-ASR model architecture. 
We also found that the current ATPC still has a performance gap compared to manually crafted pronunciation lexicons (Row 0 versus Row 6).

\begin{table}[t]
    \renewcommand{\arraystretch}{1.5}
    \centering
    \caption{ Results for contextual biasing evaluation. The left for CER (U-CER/B-CER), and the right for F1-score (Recall/Precision).}
    \label{tab:contexual biasing result}
    \begin{tabular}{l|cc}
    \hline
    & \multicolumn{2}{c} {{Aishell1 Contextual Biasing Test Set}} \\
    \hline\hline
    Hotwords number& \multicolumn{2}{c} {$N=161$} \\
    \hline
    0. C-g + Manual Lexicons  & 8.9 (7.4/15.3) & 86 (77/98)  \\
    \hline\hline
    1. Baseline & 13.8 (7.3/41.8) & 44 (28/99)  \\
    \hline
    2. C-g & 11.1 (7.4/27.2) & 72 (57/97)  \\
    \hline
    3. CPPN\cite{huang2023contextualized} & 13.9 (7.3/42.0) & 46 (30/96)  \\
    \hline
    4. ATPC (Proposed)  & 12.0 (7.3/32.4) & 68 (53/96)  \\
    \hline
    5. C-g + CPPN  & 10.4 (7.7/22.1) & 78 (66/96)  \\
    \hline
    6. C-g + ATPC (Proposed) & \textbf{10.3} (7.7/21.5)  & \textbf{80} (70/94)  \\
    \hline
    \end{tabular}
\end{table}

\section{conclusions and future work}
\label{sec:conclusion}

In this paper, we introduced ATPC, a data-driven method for automatically deriving pronunciation correlations between text symbols. Although there are gaps between the ATPC method and manually crafted pronunciation lexicons, ATPC offers an automated approach. Our visual analysis and contextual biasing experiments validate the effectiveness of the proposed ATPC on Mandarin. The contextual biasing experiments on the Aishell-1 dataset demonstrated ATPC's effectiveness as a plug-in module for E2E-ASR, yielding a relative reduction of 13.0\% in CER and 22.5\% in B-CER. While we used Mandarin with manually crafted pronunciation lexicons to assess ATPC's performance, the method's true potential lies in its application to languages or dialects where such resources are not available.

Future work could focus on the generation and application of ATPC beyond Mandarin for multiple languages or dialects, the handling of out-of-vocabulary characters or words, leveraging larger datasets to enhance ATPC's robustness, as well as advancing the standardization and continuous updating of ATPC as a public speech resource.


\bibliographystyle{IEEEtran}
\bibliography{refs}

\begin{thebibliography}{10}
\providecommand{\url}[1]{#1}
\csname url@samestyle\endcsname
\providecommand{\newblock}{\relax}
\providecommand{\bibinfo}[2]{#2}
\providecommand{\BIBentrySTDinterwordspacing}{\spaceskip=0pt\relax}
\providecommand{\BIBentryALTinterwordstretchfactor}{4}
\providecommand{\BIBentryALTinterwordspacing}{\spaceskip=\fontdimen2\font plus
\BIBentryALTinterwordstretchfactor\fontdimen3\font minus \fontdimen4\font\relax}
\providecommand{\BIBforeignlanguage}[2]{{%
\expandafter\ifx\csname l@#1\endcsname\relax
\typeout{** WARNING: IEEEtran.bst: No hyphenation pattern has been}%
\typeout{** loaded for the language `#1'. Using the pattern for}%
\typeout{** the default language instead.}%
\else
\language=\csname l@#1\endcsname
\fi
#2}}
\providecommand{\BIBdecl}{\relax}
\BIBdecl

\bibitem{intro-asr-lexi1-jstsp}
J.~Dines, J.~Yamagishi, and S.~King, ``Measuring the gap between hmm-based asr and tts,'' \emph{IEEE Journal of Selected Topics in Signal Processing}, vol.~4, no.~6, pp. 1046--1058, 2010.

\bibitem{intro-tts-lexi1-taslp}
J.~Yamagishi, B.~Usabaev, S.~King, O.~Watts, J.~Dines, J.~Tian, Y.~Guan, R.~Hu, K.~Oura, Y.-J. Wu, K.~Tokuda, R.~Karhila, and M.~Kurimo, ``Thousands of voices for hmm-based speech synthesis–analysis and application of tts systems built on various asr corpora,'' \emph{IEEE Transactions on Audio, Speech, and Language Processing}, vol.~18, no.~5, pp. 984--1004, 2010.

\bibitem{ee-kws}
R.~Yang, G.~Cheng, H.~Miao, T.~Li, P.~Zhang, and Y.~Yan, ``Keyword search using attention-based end-to-end asr and frame-synchronous phoneme alignments,'' \emph{IEEE/ACM Transactions on Audio, Speech, and Language Processing}, vol.~29, pp. 3202--3215, 2021.

\bibitem{kaiyu-modular-ee-asr}
Q.~Liu, Z.~Chen, H.~Li, M.~Huang, Y.~Lu, and K.~Yu, ``Modular end-to-end automatic speech recognition framework for acoustic-to-word model,'' \emph{IEEE/ACM Transactions on Audio, Speech, and Language Processing}, vol.~28, pp. 2174--2183, 2020.

\bibitem{davel2009pronunciation}
M.~Davel and O.~Martirosian, ``Pronunciation dictionary development in resource-scarce environments,'' 2009.

\bibitem{pronun-for-asr}
L.~Lamel and G.~Adda, ``On designing pronunciation lexicons for large vocabulary continuous speech recognition,'' in \emph{Proceeding of Fourth International Conference on Spoken Language Processing. ICSLP '96}, vol.~1, 1996, pp. 6--9 vol.1.

\bibitem{li2021recent}
J.~Li, ``Recent advances in end-to-end automatic speech recognition,'' \emph{APSIPA Transactions on Signal and Information Processing}, 2021.

\bibitem{eteh-ee-asr}
G.~Cheng, H.~Miao, R.~Yang, K.~Deng, and Y.~Yan, ``Eteh: Unified attention-based end-to-end asr and kws architecture,'' \emph{IEEE/ACM Transactions on Audio, Speech, and Language Processing}, vol.~30, pp. 1360--1373, 2022.

\bibitem{las-ee-asr}
W.~Chan, N.~Jaitly, Q.~Le, and O.~Vinyals, ``Listen, attend and spell: A neural network for large vocabulary conversational speech recognition,'' in \emph{2016 IEEE International Conference on Acoustics, Speech and Signal Processing (ICASSP)}, 2016, pp. 4960--4964.

\bibitem{online-ee-asr}
H.~Miao, G.~Cheng, P.~Zhang, and Y.~Yan, ``Online hybrid ctc/attention end-to-end automatic speech recognition architecture,'' \emph{IEEE/ACM Transactions on Audio, Speech, and Language Processing}, vol.~28, pp. 1452--1465, 2020.

\bibitem{bruguier2019phoebe}
A.~Bruguier, R.~Prabhavalkar, G.~Pundak, and T.~N. Sainath, ``Phoebe: Pronunciation-aware contextualization for end-to-end speech recognition,'' in \emph{ICASSP 2019-2019 IEEE International Conference on Acoustics, Speech and Signal Processing (ICASSP)}.\hskip 1em plus 0.5em minus 0.4em\relax IEEE, 2019, pp. 6171--6175.

\bibitem{Towards-Contextual-Spelling-Correction-ee-asr}
\BIBentryALTinterwordspacing
X.~Wang, Y.~Liu, J.~Li, V.~Miljanic, S.~Zhao, and H.~Khalil, ``Towards contextual spelling correction for customization of end-to-end speech recognition systems,'' \emph{IEEE/ACM Trans. Audio, Speech and Lang. Proc.}, vol.~30, p. 3089–3097, sep 2022. [Online]. Available: \url{https://doi.org/10.1109/TASLP.2022.3205753}
\BIBentrySTDinterwordspacing

\bibitem{itse}
R.~Yang, G.~Cheng, P.~Zhang, and Y.~Yan, ``An e2e-asr-based iteratively-trained timestamp estimator,'' \emph{IEEE Signal Processing Letters}, vol.~29, pp. 1654--1658, 2022.

\bibitem{graves2006ctc}
A.~Graves, S.~Fern{\'a}ndez, F.~Gomez, and J.~Schmidhuber, ``Connectionist temporal classification: labelling unsegmented sequence data with recurrent neural networks,'' in \emph{Proceedings of the 23rd international conference on Machine learning}, 2006, pp. 369--376.

\bibitem{pasad2021layer}
A.~Pasad, J.-C. Chou, and K.~Livescu, ``Layer-wise analysis of a self-supervised speech representation model,'' in \emph{2021 IEEE Automatic Speech Recognition and Understanding Workshop (ASRU)}.\hskip 1em plus 0.5em minus 0.4em\relax IEEE, 2021, pp. 914--921.

\bibitem{dtw}
M.~M{\"u}ller, ``Dynamic time warping,'' \emph{Information retrieval for music and motion}, pp. 69--84, 2007.

\bibitem{babel}
M.~J. Gales, K.~M. Knill, A.~Ragni, and S.~P. Rath, ``Speech recognition and keyword spotting for low-resource languages: Babel project research at cued,'' in \emph{Fourth International workshop on spoken language technologies for under-resourced languages (SLTU-2014)}.\hskip 1em plus 0.5em minus 0.4em\relax International Speech Communication Association (ISCA), 2014, pp. 16--23.

\bibitem{du2018aishell2}
J.~Du, X.~Na, X.~Liu, and H.~Bu, ``Aishell-2: Transforming mandarin asr research into industrial scale,'' \emph{arXiv preprint arXiv:1808.10583}, 2018.

\bibitem{bu2017aishell1}
H.~Bu, J.~Du, X.~Na, B.~Wu, and H.~Zheng, ``Aishell-1: An open-source mandarin speech corpus and a speech recognition baseline,'' in \emph{2017 20th conference of the oriental chapter of the international coordinating committee on speech databases and speech I/O systems and assessment (O-COCOSDA)}.\hskip 1em plus 0.5em minus 0.4em\relax IEEE, 2017, pp. 1--5.

\bibitem{huang2023contextualized}
K.~Huang, A.~Zhang, Z.~Yang, P.~Guo, B.~Mu, T.~Xu, and L.~Xie, ``Contextualized end-to-end speech recognition with contextual phrase prediction network,'' in \emph{Annual Conference of the International Speech Communication Association, INTERSPEECH 2023}, 2023, pp. 4933--4937.

\bibitem{conneau21_interspeech}
A.~Conneau, A.~Baevski, R.~Collobert, A.~Mohamed, and M.~Auli, ``Unsupervised cross-lingual representation learning for speech recognition,'' in \emph{Proc. Interspeech 2021}, 2021, pp. 2426--2430.

\bibitem{zhang2022wenet}
B.~Zhang, D.~Wu, Z.~Peng, X.~Song, Z.~Yao, H.~Lv, L.~Xie, C.~Yang, F.~Pan, and J.~Niu, ``Wenet 2.0: More productive end-to-end speech recognition toolkit,'' in \emph{Proc. Interspeech 2022}, 2022, pp. 1661--1665.

\bibitem{zhao2019shallow}
D.~Zhao, T.~N. Sainath, D.~Rybach, P.~Rondon, D.~Bhatia, B.~Li, and R.~Pang, ``Shallow-fusion end-to-end contextual biasing.'' in \emph{Proc. Interspeech 2019}, 2019, pp. 1418--1422.

\bibitem{kim2021improved-shallow-fusion}
S.~Kim, Y.~Shangguan, J.~Mahadeokar, A.~Bruguier, C.~Fuegen, M.~L. Seltzer, and D.~Le, ``Improved neural language model fusion for streaming recurrent neural network transducer,'' in \emph{ICASSP 2021-2021 IEEE International Conference on Acoustics, Speech and Signal Processing (ICASSP)}.\hskip 1em plus 0.5em minus 0.4em\relax IEEE, 2021, pp. 7333--7337.

\bibitem{han2021cif-deep-fusion}
M.~Han, L.~Dong, S.~Zhou, and B.~Xu, ``Cif-based collaborative decoding for end-to-end contextual speech recognition,'' in \emph{ICASSP 2021-2021 IEEE International Conference on Acoustics, Speech and Signal Processing (ICASSP)}.\hskip 1em plus 0.5em minus 0.4em\relax IEEE, 2021, pp. 6528--6532.

\bibitem{chen2019joint-deep-fusion}
Z.~Chen, M.~Jain, Y.~Wang, M.~L. Seltzer, and C.~Fuegen, ``Joint grapheme and phoneme embeddings for contextual end-to-end asr.'' in \emph{Proc. Interspeech 2019}, 2019, pp. 3490--3494.

\end{thebibliography}
\end{CJK}
\end{document}